# Closures in Formal Languages: Concatenation, Separation, and Algorithms

Janusz Brzozowski, Elyot Grant, and Jeffrey Shallit

David R. Cheriton School of Computer Science,
University of Waterloo, Waterloo, Ontario, Canada N2L 3G1
{brzozo, egrant, shallit}@cs.uwaterloo.ca

**Abstract.** We continue our study of open and closed languages. We investigate how the properties of being open and closed are preserved under concatenation. We investigate analogues, in formal languages, of the separation axioms in topological spaces; one of our main results is that there is a clopen partition separating two words if and only if the words commute. We show that we can decide in quadratic time if the language specified by a DFA is closed, but if the language is specified by an NFA, the problem is PSPACE-complete.

## 1 Introduction

In a previous paper [2], we extended the work of Peleg [6] on closure operators of formal languages. In that paper, we observed that both positive and Kleene closure can be viewed as instances of a closure operator $\square$; a language $L$ is *closed* if $L = L^{\square}$. Similarly, a language is *open* if it is the complement of a closed language. A language is *clopen* if it is both closed and open. We proved many properties of open and closed languages, which share some but not all properties with analogous concepts in topological spaces. We proved two versions of Kuratowski's theorem for applying any number of the operators of closure and complement in any order, and we gave a complete characterization of all algebras resulting from this process.

In this paper, we continue our study of open and closed languages. In Section 2, we investigate how the properties of being open and closed are preserved under concatenation. In Section 3, we investigate analogues, in formal languages, of the separation axioms in topological spaces; one of our main results (Theorem 6) is that there is a clopen partition separating two words if and only if the words commute. In Section 4, we show that we can decide in quadratic time if the language specified by a DFA is closed, but if the language is specified by an NFA, the problem is PSPACE-complete. Finally, in Section 5, we mention some analogues of the compactness property.

## 2 Closure operators and concatenation

We recall two results from [2]:

**Proposition 1.** *Let $L \subseteq \Sigma^*$.*

*(a) $L$ is positive-closed if and only if, for all $u, v \in L$, we have $uv \in L$.*

*(b) $L$ is positive-open if and only if for all $u, v \in \Sigma^*$ such that $uv \in L$, we have $u \in L$ or $v \in L$.*

We note that the concatenation of two closed languages need not be closed, and that the concatenation of two open languages need not be open. Consider the languages $L = \{a\}^+$ and $M = \{b\}^+$ for $a, b \in \Sigma$, which are both clopen (under positive closure). Then $ab \in LM$ but $abab \notin LM$, so $LM$ is not closed. Additionally, $ab \in LM$, but neither $a$ nor $b$ is in $LM$, so $LM$ is not open. However, we do have several results regarding cases when the concatenation of closed or open languages must be closed or open.

Throughout this paper, the words *closed*, *open*, and *clopen* refer to their respective notions under *positive* closure, as it is the most general—most of our theorems will have obvious analogues in the Kleene closure case. However, the presence of $\epsilon$ (or lack thereof) can be crucial when dealing with the concatenation of languages, so we will mention a few exceptional cases where the choice of positive or Kleene closure is important.

**Theorem 1.** *Let $L, M \subseteq \Sigma^*$.*

*(a) Suppose $L$ is positive-closed, and let $k$ be a positive integer. Then $L^k \subseteq L$ and $L^k$ is positive-closed.*

*(b) Suppose $L$ is Kleene-closed, and let $k$ be a positive integer. Then $L^k = L$.*

*(c) Suppose $L$ and $M$ are positive-closed (respectively, Kleene-closed) languages satisfying the equation $LM = ML$. Then $LM$ is positive-closed (respectively, Kleene-closed).*

*(d) Suppose $L$ and $M$ are positive-closed (respectively, Kleene-closed) unary languages (that is, $L, M \subseteq \{a\}^*$ for some symbol $a$.) Then $LM$ is positive-closed (respectively, Kleene-closed).*

*Proof.* (a) Let $u, v \in L^k$. We show that $L^k$ is positive-closed by proving that $uv \in L^k$. Let $u = u_1 u_2 \cdots u_k$ and $v = v_1 v_2 \cdots v_k$ where $u_1, \ldots, u_k, v_1, \ldots, v_k \in L$. Then $u \in L$ since $L$ is closed and thus $L^k \subseteq L$. Similarly, $uv_1 \in L^{k+1} \subseteq L$, and thus $uv = (uv_1)v_2 \cdots v_k \in L^k$. So $L^k$ is positive-closed.

(b) If $L$ is Kleene-closed, then $L = L\epsilon^{k-1} \subseteq L^k$. But $L$ must also be positive-closed, so $L^k \subseteq L$ by part (a). Thus $L^k = L$.

(c) We examine the positive-closed case first. If $u, v \in LM$, then let $u = u_1 u_2$ and $v = v_1 v_2$ where $u_1, v_1 \in L$ and $u_2, v_2 \in M$. Then $u_2 v_1 \in LM$ since $L$ and $M$ commute. Thus $uv = u_1 u_2 v_1 v_2 \in LLMM$. But $LL \subseteq L$ and $MM \subseteq M$ by part (a), so $LLMM \subseteq LM$, implying that $uv \in LM$ and thus $LM$ is positive-closed. For the Kleene-closed case, we simply note that if $\epsilon \in L$ and $\epsilon \in M$ then $\epsilon \in LM$ and the result follows.

(d) This is a special case of part (c), since unary languages commute.  ∎

**Theorem 2.** *Let $L, M \subseteq \Sigma^*$. Suppose $L$ and $M$ are positive-closed (respectively, Kleene-closed) and such that $L \cup M$ is positive-closed. Then the following hold:*

(a) $LM$ is positive-closed (respectively, Kleene-closed).
(b) More generally, consider the free semigroup of languages $\{L, M\}^+$ generated by $L$ and $M$. Let $W \in \{L, M\}^+$. Then $W$ is positive-closed (respectively, Kleene-closed) when considered as a language over $\Sigma$.

*Proof.* (a) Let $u, v \in LM$, then let $u = u_1 u_2$ and $v = v_1 v_2$ where $u_1, v_1 \in L$ and $u_2, v_2 \in M$. If $L \cup M$ is positive-closed, then $L \cup M = (L \cup M)^+$. But $u_2 v_1 \in (L \cup M)^+ = L \cup M$, so either $u_2 v_1 \in L$ or $u_2 v_1 \in M$. If $u_2 v_1 \in L$, then $u_1 u_2 v_1 \in L$ by closure of $L$, and thus $u_1 u_2 v_1 v_2 \in LM$. Similarly, if $u_2 v_1 \in M$, then $u_2 v_1 v_2 \in M$, and hence $u_1 u_2 v_1 v_2 \in LM$. So in all cases we have $uv = u_1 u_2 v_1 v_2 \in LM$ and hence $LM$ is closed. For the Kleene-closed case, we again simply note that if $\epsilon \in L$ and $\epsilon \in M$, then $\epsilon \in LM$.

(b) The cases where $W = L^k$ or $W = M^k$ are proven by Theorem 1, so we may assume that $W$ contains at least one $L$ and one $M$ (when considered as a word in $\{L, M\}^+$.) This implies that either $LM$ or $ML$ is a factor of $W$. Suppose without loss of generality that $LM$ is a factor of $W$. Let $W = W_1 W_2 \cdots W_k W_{k+1} \cdots W_n$ where $W_i \in \{L, M\}$ for all $i$, and specifically $W_k = L$ and $W_{k+1} = M$. Now, to prove that $W$ is closed, we let $u, v \in W$ be words in $\Sigma^*$. We show that $uv \in W$. Let $u = u_1 \cdots u_n$ and $v = v_1 \cdots v_n$ where $u_i, v_i \in W_i$ for all $i$. Consider $x = u_{k+1} \cdots u_n v_1 \cdots v_k$, a factor of $uv$. We see that $x \in (L \cup M)^+$. But since $L \cup M$ is positive-closed, $(L \cup M)^+ = L \cup M$, and hence either $x \in L$ or $x \in M$. If $x \in L$, then $u_k x \in L = W_k$ by closure of $L$ and thus $uv = u_1 \cdots u_{k-1} (u_k x) v_{k+1} \cdots v_n \in W_1 \cdots W_n = W$. If $x \in M$, then $xv_{k+1} \in M = W_{k+1}$ by closure of $M$ and thus $uv = u_1 \cdots u_k (xv_{k+1}) v_{k+2} \cdots v_n \in W_1 \cdots W_n = W$. So we must have $uv \in W$ in either case, and thus $W$ is closed. For the Kleene-closed case, we again simply note that if $\epsilon \in L$ and $\epsilon \in M$, then $\epsilon \in W$. ∎

**Theorem 3.** *Let $L$ and $M$ be open.*

(a) Suppose $\epsilon \in L$ and $\epsilon \in M$. Then $LM$ is open.
(b) Suppose $\epsilon \notin L$ and $\epsilon \notin M$. Then $LM$ is open if and only if $L = \emptyset$ or $M = \emptyset$.
(c) $LL$ is open if and only if $\epsilon \in L$ or $L = \emptyset$.
(d) If neither $L$ nor $M$ is empty and $\epsilon \in L \cup M$ but $\epsilon \notin L \cap M$, then we may or may not have $LM$ open, even in the unary case.

*Proof.* (a) Let $ab \in LM$ where $a \in L$ and $b \in M$. Let $ab = uv$ for some words $u$ and $v$. To prove that $LM$ is open, we must show that either $u \in LM$ or $v \in LM$. We have two cases: either $u$ is a prefix of $a$, or $v$ is a suffix of $b$.

If $u$ is a prefix of $a$, let $a = ux$, so $ab = uxb$ and hence $v = xb$. Since $L$ is open, applying Proposition 1 (b) to $a \in L$ implies that either $u \in L$ or $x \in L$. If $u \in L$, then since $\epsilon \in M$, $u = u\epsilon \in LM$ and we are done. If $x \in L$, then $v = xb \in LM$ and we are also done.

The case where $v$ is a prefix of $b$ is similar and relies on us having $\epsilon \in L$.

(b) If $L = \emptyset$ or $M = \emptyset$, then $LM = \emptyset$, which is open. Conversely, if $\epsilon \notin L$, $\epsilon \notin M$, and neither $L = \emptyset$ nor $M = \emptyset$, then $LM$ is non-empty but contains no words of length 0 or 1 and is thus not open.

(c) This follows immediately from parts (a) and (b).
(d) If $L = \{\epsilon, a, aaa, aaaaa\}$ and $M = \{a\}$ (which are both easily verified to be open), then we have $aaaaaa \in LM$, but $aaa \notin LM$, and thus $LM$ is not open. On the other hand, if $L = \{\epsilon, a, aaa\}$ and $M = \{a\}$, then $LM = \{a, aa, aaaa\}$ which is clearly open.

∎

**Theorem 4.** *Let $L, M \subseteq \Sigma^*$ both be clopen.*

(a) *If $L \cup M = \Sigma^*$, then $LM$ is clopen.*
(b) *Suppose that $L \cup M = \Sigma^*$ and consider the free semigroup of languages $\{L, M\}^+$ generated by $L$ and $M$. Let $W \in \{L, M\}^+$. Then $W$ is clopen if and only if $W = \emptyset$ or $W$ contains at most one occurrence of a language which does not contain $\epsilon$.*
(c) *The converses of the above statements are false; indeed, it is possible that $LM$ is clopen, but $L \cup M$ is not even positive-closed.*

*Proof.* (a) From Theorem 2 (a) we have that $LM$ is closed, since $\Sigma^*$ is closed. To show that $LM$ is open, let $ab \in LM$ where $a \in L$ and $b \in M$. Let $ab = uv$ for some words $u$ and $v$. To prove that $LM$ is open, we must show that either $u \in LM$ or $v \in LM$. There are two cases: either $u$ is a prefix of $a$, or $v$ is a suffix of $b$.

Without loss of generality, we assume that $u$ is a prefix of $a$ and let $a = ux$, so $ab = uxb$ and hence $v = xb$. Since $L$ is open, applying Proposition 1 (b) to $a \in L$ implies that either $u \in L$ or $x \in L$. If $x \in L$, then $v = xb \in LM$ and we are done. Otherwise, we have $x \notin L$, implying $u \in L$ and $x \in M$ since $L \cup M = \Sigma^*$. If $\epsilon \in M$, $u = u\epsilon \in LM$ and we are done. Otherwise, we have $\epsilon \notin M$, and thus $\epsilon \in L$ since $L \cup M = \Sigma^*$. In this case, we note that $xb \in M$ since $x \in M$, $b \in M$, and $M$ is closed. Then $\epsilon xb = v \in LM$. So in all cases, we have either $u \in LM$ or $v \in LM$. Thus $LM$ is open and hence is clopen.

(b) Let $W = W_1 W_2 \cdots W_n$ where $W_i \in \{L, M\}$ for all $i$. $W$ is closed by Theorem 2 (b). If each $W_i$ contains $\epsilon$, then $W$ is open by repeated applications of Theorem 3 (a) and is thus clopen.

If there exist $i$ and $j$ with $i \neq j$, $\epsilon \notin W_i$, and $\epsilon \notin W_j$, then $W$ contains no words of length 1, so either $W = \emptyset$ or $W$ is not open (and thus not clopen).

Finally, we deal with the case where there exists a unique $i$ such that $\epsilon \notin W_i$. Suppose, without loss of generality, that $W_i = M$. Then $W = L^{i-1} M L^{n-i}$. Since $L \cup M$ is Kleene-closed, it must contain $\epsilon$, so $\epsilon \in L$. Thus $L^k = L$ for all positive $k$ by Theorem 1 (b), so we must have $W = M$, $W = LM$, $W = ML$, or $W = LML$. In the first case, $W = M$ is known to be clopen, and in the second and third cases, $W$ is clopen by part (a). Thus we must only consider the case where $W = LML$. We know that $LM$ is clopen by part (a). Furthermore, $M \subseteq LM$ since $\epsilon \in L$, so $LM \cup L \supseteq M \cup L = \Sigma^*$ and thus $LM \cup L = \Sigma^*$. Thus we can apply part (a) on $LM$ and $L$, proving that $LML$ is clopen.

(c) As a counterexample, we let $L = \{\epsilon\} \cup \{w \in \{a,b\}^* : |w|_a < |w|_b\}$ and let $M = \{\epsilon\} \cup \{w \in \{a,b\}^* : |w|_a > |w|_b\}$, where by $|w|_c$ for a letter $c$, we mean the number of occurrences of $c$ in $w$. As we proved in [2, Example 1], $L$ and $M$ are both clopen. Furthermore, $L$ and $M$ both contain $\epsilon$, so $LM$ is open by Theorem 3.

Next, we show that $LM$ is closed. Let $u, v \in LM$, then let $u = u_1 u_2$ and $v = v_1 v_2$, where $u_1, v_1 \in L$ and $u_2, v_2 \in M$. We observe that $|u_1|_a < |u_1|_b$ and $|v_2|_a > |v_2|_b$. We examine the factor $u_2 v_1$ and consider two cases. If $|u_2 v_1|_a \geq |u_2 v_1|_b$, then $|u_2 v_1 v_2|_a > |u_2 v_1 v_2|_b$ and thus $u_2 v_1 v_2 \in M$. Since $u_1 \in L$, we must then have $uv = u_1 u_2 v_1 v_2 \in LM$. Similarly, if $|u_2 v_1|_a \leq |u_2 v_1|_b$, then $|u_1 u_2 v_1|_a < |u_1 u_2 v_1|_b$ and thus $u_1 u_2 v_1 \in L$. Since $v_2 \in M$, we must then have $uv = u_1 u_2 v_1 v_2 \in LM$. So in all cases, $uv \in LM$, and $LM$ is closed. Hence $LM$ is clopen.

However, $L \cup M$ is not closed, since we have $b \in L \subseteq L \cup M$ and $a \in M \subseteq L \cup M$, but $ba \notin L \cup M$.

∎

## 3 Separation of words and languages

Next, we discuss analogies of the separation axioms of topology in the realm of languages. Although languages do not form a topology under Kleene or positive closure, there are many interesting results describing when there exist open, closed, and clopen languages that separate given words or languages. In most of these theorems, we only consider words in $\Sigma^+$, as $\epsilon$ is always a trivial case.

**Lemma 1.** *Let $w \in \Sigma^+$, and let $L \subseteq \Sigma^*$ be closed with $w \notin L$. Then there exists a finite open language $M$ such that $w \in M$ but $M \cap L = \emptyset$,*

*Proof.* We simply take $M = L^- \cap \{x \in \Sigma^+ : |x| \leq |w|\}$. This is clearly finite, and is open by our characterization. ∎

**Theorem 5.** *Let $u, v \in \Sigma^+$.*

(a) *There exists an open language $L$ with $u \in L$ and $v \notin L$ if and only if for all natural numbers $k$, we have $u \neq v^k$.*
(b) *If $u \neq v$, then either there exists an open language $L$ with $u \in L$ and $v \notin L$, or there exists an open language $L$ with $u \notin L$ and $v \in L$ (all words are distinguishable by open languages).*

*Proof.* (a) For the forward direction, we note that if $u = v^k$ for some positive $k$, then any open language containing $u$ must contain $v$ by Proposition 1 (b). For the reverse direction, we apply Lemma 1 to $u$ and $\{v\}^+$, which is closed.

(b) If $u \neq v$, we can just pick $w \in \{u, v\}$ such that $|w|$ in minimal. Then we take $L = \text{Pref}(\{w\})$, which must contain one of $u$ or $v$ but not the other, and is open as we saw in [2, Example 2]. Here $\text{Pref}(L)$ is the language of all prefixes of words in $L$.

∎

We now recall a basic result from combinatorics on words (see, e.g., [5]). Recall that a word $w$ is *primitive* if it cannot be expressed in the form $x^k$ for a word $x$ and an integer $k \geq 2$.

**Lemma 2.** *Let $u, v \in \Sigma^+$. The following are equivalent:*

*(1) $uv = vu$, that is, $u$ and $v$ commute.*
*(2) There exists a word $x$ and integers $p \geq 1$ and $q \geq 1$ such that $u = x^p$ and $v = x^q$.*
*(3) There exists a word $y$ and integers $p \geq 1$ and $q \geq 1$ such that $y = u^p$ and $y = v^q$.*
*(4) $u$ and $v$ are each a power of the same primitive word.*

Let $u, v \in \Sigma^+$. Suppose there exists a clopen language $L \subseteq \Sigma^*$ with $u \in L$ and $v \notin L$. We note that $L^-$ is also clopen whenever $L$ is, and we call the pair $(L, L^-)$ a *clopen partition separating $u$ and $v$*.

**Theorem 6.** *Let $u, v \in \Sigma^+$. There exists a clopen partition separating $u$ and $v$ if and only if $u$ and $v$ do not commute.*

*Proof.* We handle the forward direction first. Suppose a clopen language $L$ exists with $u \in L$ and $v \notin L$. If $u$ and $v$ commute, then there exists a word $x$ and integers $p$ and $q$ such that $u = x^p$ and $v = x^q$. In particular, this implies that any open set containing $u$ will also contain $x$, and any open set containing $v$ will also contain $x$. Then we must have both $x \in L$ (since $L$ is open and contains $u$) and $x \in L^-$, since $L^-$ is open and contains $v$. Thus we have a contradiction, and $u$ and $v$ must not commute.

For the reverse direction, we proceed by induction on $|u| + |v|$. We will apply the induction hypothesis on words in various alphabets, so we make no assumption that $|\Sigma|$ is constant.

For our base case, suppose $|u| + |v| = 2$. If $u$ and $v$ do not commute, then they must be distinct words of length 1, and thus the language $\{u\}^+$ is a clopen language separating $u$ from $v$.

Suppose, as a hypothesis, that for some $k \geq 2$, the result holds for all finite alphabets $\Sigma$ and for all $u, v \in \Sigma^+$ such that $2 \leq |u| + |v| \leq k$. Now, given any $\Sigma$, let $u, v \in \Sigma^+$ be such that $u$ and $v$ do not commute and $|u| + |v| = k + 1$. Let $\Sigma_u$ and $\Sigma_v$, respectively, be the symbols that occur one or more times in $u$ and $v$. If $\Sigma_u \cap \Sigma_v = \emptyset$, then $\Sigma_u^+$ is a clopen language containing $u$ but not $v$, and our result holds. If not, suppose $a \in \Sigma_u \cap \Sigma_v$. Let $\lambda_u = \frac{|u|_a}{|u|}$ and $\lambda_v = \frac{|v|_a}{|v|}$ be the respective relative frequencies of $a$ in $u$ and $v$. If $\lambda_u > \lambda_v$, then $\{w \in \Sigma^* : |w|_a \geq \lambda_u |w|\}$ is clopen (by [2, Example 1]) and contains $u$ but not $v$, and we are done. Similarly, if $\lambda_u < \lambda_v$, then $\{w \in \Sigma^* : |w|_a \leq \lambda_u |w|\}$ is a clopen language containing $u$ but not $v$. Thus it remains to show that the result holds when $\lambda_u = \lambda_v$.

Assume $\lambda_u = \lambda_v = \lambda$. If $\lambda = 1$, then $u = a^i$ and $v = a^j$ for some positive integers $i$ and $j$, and thus $u$ and $v$ commute, contradicting our original assumption. Hence we must have $0 < \lambda < 1$. Let $n = \frac{|u|}{\gcd(|u|_a, |u|)} = \frac{|v|}{\gcd(|v|_a, |v|)}$ be the denominator of $\lambda$ when it is expressed in lowest terms. We must have $n > 1$ since $\lambda$ is not an integer.

Next, we consider a new alphabet $\Delta$ with $|\Sigma|^n$ symbols, each corresponding to a word of length $n$ in $\Sigma^*$. We consider the bijective morphism $\phi$ mapping words in $\Delta^*$ to words in $(\Sigma^n)^*$ by replacing each symbol in $\Delta$ with its corresponding word in $\Sigma^n$. Since $n$ divides both $|u|$ and $|v|$, there must then exist unique words $p, q \in \Delta^*$ such that $\phi(p) = u$ and $\phi(q) = v$.

Our plan is now to inductively create a clopen language $L$ over $\Delta$ which contains $p$ but not $q$, and then use this language to construct our clopen partition over $\Sigma$ separating $u$ and $v$. We must check that $p$ and $q$ do not commute. If $pq = qp$ then we would have $uv = \phi(p)\phi(q) = \phi(pq) = \phi(qp) = \phi(q)\phi(p) = vu$, since $\phi$ is a morphism. This is impossible since $uv \neq vu$, so $p$ and $q$ do not commute. We also have $n|p| + n|q| = |u| + |v|$. Since $n > 1$ implies $|p| + |q| < |u| + |v| = k+1$, the induction hypothesis can be applied to $p$ and $q$. Thus there exists a clopen language $L \subseteq \Delta^*$ with $p \in L$ and $q \notin L$.

We now construct our clopen partition over $\Sigma$ separating $u$ and $v$. We introduce some notation to make this easier. As usual, define $\phi(L) = \{w \in \Sigma^* : w = \phi(r) \text{ for some } r \in L\}$. Let $A^< = \{w \in \Sigma^* : |w|_a < \lambda|w|\}$ and let $A^= = \{w \in \Sigma^* : |w|_a = \lambda|w|\}$. Additionally, let $A^\leq = A^< \cup A^=$. It is easy to verify that $A^<$, $A^\leq$, and $A^=$ are all closed, and both $A^<$ and $A^\leq$ are open as well. Finally, we let $M = (\phi(L) \cap A^=) \cup A^<$. Since $p \in L$ and $q \notin L$, we must have $u \in \phi(L)$ and $v \notin \phi(L)$. Then since $u$ and $v$ are both contained in $A^=$ but not $A^<$, we must have $u \in M$ and $v \notin M$. We will now finish the proof by showing that $M$ is clopen.

We first show that $M$ is closed. Let $x, y \in M$. We must show that $xy \in M$. There are two cases to consider:

Case (A1): $x, y \in (\phi(L) \cap A^=)$. We see that $\phi(L)\phi(L) = \phi(LL) \subseteq \phi(L)$, so $\phi(L)$ is closed. Then since $A^=$ is closed, $(\phi(L) \cap A^=)$ is the intersection of two closed languages, and hence closed. Thus $xy \in (\phi(L) \cap A^=) \subseteq M$.

Case (A2): One or more of $x$ or $y$ is not in $(\phi(L) \cap A^=)$. Without loss of generality, suppose $x \notin (\phi(L) \cap A^=)$. Then $x \in A^<$, so $|x|_a < \lambda|x|$. Furthermore, $y \in M \subseteq A^\leq$, so $|y|_a \leq \lambda|y|$. Adding these two inequalities yields $|x|_a + |y|_a < \lambda|x| + \lambda|y|$, so $|xy|_a < \lambda|xy|$ and thus $xy \in A^< \subseteq M$.

Lastly, we show that $M$ is open. Let $z \in M$ and suppose $z = xy$ for some $x, y \in \Sigma^+$. We show that $x \in M$ or $y \in M$. Again, we have two cases to consider:

Case (B1): $z \in A^<$. Since $A^<$ is open, at least one of $x$ or $y$ is in $A^<$. Since $A^< \subseteq M$, we are done.

Case (B2): $z \in (\phi(L) \cap A^=)$. If either $x$ or $y$ is in $A^<$, then we are done, so assume otherwise. Then $|x|_a \geq \lambda|x|$ and $|y|_a \geq \lambda|y|$. But $|xy|_a = \lambda|xy|$, so we must have $|x|_a = \lambda|x|$ and $|y|_a = \lambda|y|$ and thus $x, y \in A^=$. Then $\lambda|x|$ and $\lambda|y|$ must be integers and hence $n$ divides both $|x|$ and $|y|$. Then there exist $s, t \in \Delta^*$ such that $\phi(s) = x$ and $\phi(t) = y$. But since $\phi$ is a morphism, we must then have $\phi(st) = \phi(s)\phi(t) = xy = z$. But $z$ in $\phi(L)$, so $st \in L$. Since $L$ is open, we must then have either $s \in L$ or $t \in L$. Thus we must have either $x = \phi(s) \in \phi(L)$ or $y = \phi(t) \in \phi(L)$. Then one of $x$ or $y$ is in $\phi(L) \cap A^= \subseteq M$.

Thus $M$ is both closed and open, and the result follows by induction. ∎

**Corollary 1.** *Let $u, v \in \Sigma^+$. There exist non-intersecting finite open languages $L$ and $M$ with $u \in L$ and $v \in M$ if and only if $u$ and $v$ do not commute.*

*Proof.* As in the proof of Theorem 6, we note that if $u$ and $v$ commute, then there is some $x$ such that $u = x^p$ and $v = x^q$, implying that every open language containing $u$ or $v$ must contain $x$, and thus there is no open language containing $u$ but not $v$. If $u$ and $v$ do not commute, then by our theorem, let $K$ be a clopen language containing $u$ but not $v$. We then take $L = \{w \in K : |w| \leq |u|\}$ and $M = \{w \in K^- : |w| \leq |v|\}$. These are open by our Proposition 1 (b) since $K$ and $K^-$ are both open. ∎

We can also use Theorem 6 to extend the topological notion of *connected components* to the setting of formal languages. We say that words $u, v \in \Sigma^+$ are *disconnected* if there exists a clopen partition separating $u$ from $v$, and *connected* otherwise. We write $u \sim v$ if $u$ and $v$ are connected, and note that $\sim$ is an equivalence relation (indeed, this is the case when we consider the clopen partitions created by any closure operator; it need not be topological). Since Theorem 6 implies that $u \sim v$ if and only if $u = x^p$ and $v = x^q$ for some integers $p$ and $q$, it follows that each connected component of $\Sigma^+$ consists of a primitive word and all of its powers. Connected components of other languages will simply consist of collections of words sharing a common primitive root.

It should be noted that connected components must be closed, but they need not be clopen. In fact, the only clopen components of $\Sigma^+$ are the languages $\{a\}^+$ for each $a \in \Sigma$.

As in [2], we say that a closure operator $\square$ *preserves openness* if $L^\square$ is open for all open sets $L$. We recall that positive closure preserves openness [2, Theorem 3], and use it to prove the following theorem, which indeed holds for all closure operators that preserve opennness.

**Theorem 7.** *If $L, M \subseteq \Sigma^*$ are disjoint and open, then $L^+$ and $M^+$ are disjoint.*

*Proof.* If $L \cap M = \emptyset$, then $M \subseteq L^-$. Then by isotonicity, $M^+ \subseteq L^{-+} = L^-$ since $L^-$ is closed. But then $L \subseteq M^{+-}$. Applying isotonicity again yields $L^+ \subseteq M^{+-+}$. But $M^+$ is the closure of an open language and is thus clopen, so $M^{+-}$ is also clopen and thus $M^{+-+} = M^{+-}$. Hence $L^+ \subseteq M^{+-}$, and it follows that $L^+$ and $M^+$ are disjoint. ∎

**Corollary 2.** *Let $L, M \subseteq \Sigma^*$ be closed and such that $L \cup M = \Sigma^*$. Then $L^\oplus \cup M^\oplus = \Sigma^*$.*

In our setting, it is not true that a single "point" $x$ and a closed set $S$ can be separated by two open sets. As a counterexample, consider $x = ab$ and $y = \{aa, bb\}^*$. Furthermore, it is not true that that arbitrary disjoint sets, even ones whose closures are disjoint, can be clopen separated. As an example, consider $\{ab\}^*$ and $\{aa, bb\}^*$.

## 4 Algorithms

We now consider the computational complexity of determining if a given language $L$ is closed or open. Of course, the answer depends on how $L$ is represented.

**Theorem 8.** *Given an n-state DFA $M = (Q, \Sigma, \delta, q_0, F)$ accepting the regular language $L$, we can determine in $O(n^2)$ time if $L$ is closed or open.*

*Proof.* We prove the result when $L$ is positive-closed. For Kleene-closed, we have the additional check $q_0 \in F$. For the open case, we start with a DFA for $\overline{L}$.

We know from Proposition 1 (a) that $L$ is closed if and only if, for all $u, v \in L$ we have $uv \in L$. Given $M$, we create an NFA-$\epsilon$ $M'$ that accepts all words $x \notin L$ such that there exists a decomposition $x = uv$ with $u, v \in L$. Then $L(M')$ is empty if and only if $L$ is closed.

Here is the construction of $M'$: $M' = (Q', \Sigma, \delta', q'_0, F')$, where $Q' = Q \cup Q \times Q$, $q'_0 = q_0$, $F' = (Q - F) \times F$, and $\delta'$ is defined as follows:

$$\delta'(p, a) = \{\delta(p, a)\} \text{ for } p \in Q, a \in \Sigma;$$
$$\delta'(p, \epsilon) = \{[p, q_0]\}, \text{ if } p \in F;$$
$$\delta'([p, q], a) = \{[\delta(p, a), \delta(q, a)]\} \text{ for } p, q \in Q, a \in \Sigma.$$

$M'$ functions as follows: on input $u$, it simulates the computation of $M$. If and only if a final state is reached (and so $u \in L$), $M'$ has the option to use its $\epsilon$-transition to enter a state specified by two components, the second of which is $q_0$. Now $M'$ processes $v$, determining $\delta(q_0, uv)$ in its first component and $\delta(q_0, v)$ in the second. If $uv \notin L$, but $v \in L$, then $M'$ accepts. Thus $M'$ accepts $uv$ if and only if $u, v \in L$ and $uv \notin L$.

We now use the usual depth-first search technique to determine if $L(M')$ is empty, which uses time proportional to the number of states and transitions of $M'$. Since $M'$ has $|Q||\Sigma| + |F| + |Q|^2|\Sigma|$ transitions and $|Q| + |Q|^2$ states, our depth-first search can be done in $O(n^2)$ time. ∎

From Proposition 1 (a), we know that $L$ is not closed if and only if there exists a word $uv \notin L$ such that $u, v \in L$. We call such a word a *counterexample*.

**Corollary 3.** *If $L$ is a regular language, accepted by a n-state DFA, that is not closed, then the smallest counterexample is of length $\leq n^2 + n - 1$.*

This $O(n^2)$ upper bound on the length of the shortest counterexample is matched by a corresponding $\Omega(n^2)$ lower bound:

**Theorem 9.** *There exists a class of DFA's $M_n$ with $2n + 5$ states, having the following property: the shortest word $x \notin L(M_n)$ such that there exist $u, v \in L(M_n)$ with $x = uv$ is of length $n^2 + 2n + 2$.*

*Proof.* It is conceptually easier to describe DFA's $M'_n = (Q, \Sigma, \delta, q_0, F)$ that accepts the complement of $L(M_n)$. In other words, we will show that the shortest word $x \in L(M'_n)$ such that there exist $u, v \notin L(M_n)$ with $x = uv$ is of length $n^2 + 2n + 2$. The parts of the DFA $M'_n$ are as follows:

$$Q = \{q_0, q_1, \ldots, q_n, r, p_0, p_1, \ldots, p_n, s, d\}$$
$$F = \{q_0, q_1, \ldots, q_n, p_0, p_1, \ldots, p_n, s\}$$

and $\delta$ is given in Table 1.

| $a\backslash q$ | $q_0$ | $q_1$ | $q_2$ | $\ldots$ | $q_{n-1}$ | $q_n$ | $r$ | $p_0$ | $p_1$ | $\ldots$ | $p_{n-1}$ | $p_n$ | $s$ | $d$ |
|---|---|---|---|---|---|---|---|---|---|---|---|---|---|---|
| 0 | $d$ | $q_2$ | $q_3$ | $\ldots$ | $q_n$ | $q_1$ | $d$ | $p_1$ | $p_2$ | $\ldots$ | $p_n$ | $p_0$ | $d$ | $d$ |
| 1 | $q_1$ | $s$ | $s$ | $\ldots$ | $s$ | $r$ | $p_0$ | $d$ | $d$ | $\ldots$ | $d$ | $s$ | $d$ | $d$ |

**Table 1.** Transition function $\delta(q, a)$ of $M'_n$.

The case $n = 5$ is illustrated in Figure 1.

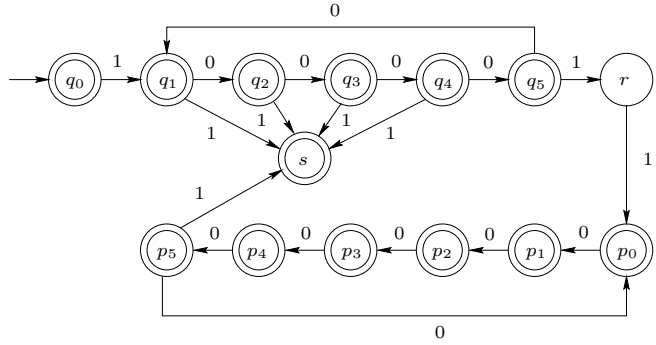

**Fig. 1.** Example of DFA $M_n$ for $n = 5$. Unspecified transitions go to the dead state $d$.

First, we observe that $x = 10^{n-1}110^{n^2+n-1}1$ is accepted by $M'_n$, but neither $u = 10^{n-1}1$ nor $v = 10^{n^2+n-1}1$ is. Next, take any word $x'$ accepted by $M'_n$. If the acceptance path does not pass through $r$, then by examining the DFA we see that every prefix of $x'$ is also accepted. Otherwise, the acceptance path passes through $r$. Again, we see that every prefix of $x'$ is accepted, with the possible exception of the prefix ending at $r$. Thus either $x'$ is of the form $10^{in+n-1}110^k$ for some $i, k \geq 0$, or $x'$ is of the form $10^{in+n-1}110^{j(n+1)+n}1$ for some $i, j \geq 0$. In both cases the prefix ending at $r$ is $10^{in+n-1}1$, so in the first case, the corresponding suffix is $10^k$ for some $k \geq 0$, and this suffix is accepted by $M'_n$. In the latter case the corresponding suffix is $10^{j(n+1)+n}1$. This is accepted unless $j(n+1) + n$ is of the form $in + n - 1$. If $in + n - 1 = j(n+1) + n$, then by taking both sides modulo $n$, we see that $j \equiv -1 \pmod{n}$. Thus $j \geq n - 1$. Thus $|x'| \geq 1 + n - 1 + 1 + 1 + (n-1)(n+1) + n + 1 = n^2 + 2n + 2$. ∎

We now turn to the case where $M$ is represented as an NFA or regular expression. We need the following classical lemma [1]:

**Lemma 3.** *Let $T$ be a one-tape deterministic Turing machine and $p(n)$ a polynomial such that $T$ never uses more than $p(|x|)$ space on input $x$. Then there is a finite alphabet $\Delta$ and a polynomial $q(n)$ such that we can construct a regular expression $r_x$ in $q(|x|)$ steps, such that $L(r_x) = \Delta^*$ if $T$ doesn't accept $x$, and $L(r_x) = \Delta^* \setminus \{w\}$ for some nonempty $w$ (depending on $x$) otherwise. Similarly, we can construct an NFA $M_x$ in $q(|x|)$ steps, such that $L(M_x) = \Delta^*$ if $T$ doesn't accept $x$, and $L(M_x) = \Delta^* \setminus \{w\}$ for some nonempty $w$ (depending on $x$) otherwise.*

For the following theorem, we actually require the word $w$ exhibited in the theorem above to have length $\geq 2$. However, this can easily be accomplished via a trivial modification of the proof given in [1], since the word $w$ encodes a configuration of the Turing machine $T$.

**Theorem 10.** *The following problem is PSPACE-complete: given an NFA $M$, decide if $L(M)$ is closed.*

*Proof.* First, we observe that the problem is in PSPACE. We give a nondeterministic polynomial-space algorithm to decide if $L(M)$ is not closed, and use Savitch's theorem to conclude the result.

If $M$ has $n$ states, then there is an equivalent DFA $M'$ with $N \leq 2^n$ states. From Corollary 3 we know that if $L = L(M) = L(M')$ is not closed, then there exist words $u, v$ with $u, v \in L$ but $uv \notin L$, and $|uv| \leq N^2 + N - 1 = 2^{2n} + 2^n - 1$. We now guess $u$, processing it symbol-by-symbol, arriving in a set of states $S$ of $M$. Next, we guess $v$, processing it symbol-by-symbol starting from both $q_0$ and $S$, respectively and ending in sets of states $T$ and $U$. If $U$ contains a state of $F$ and $T$ does not, then we have found $u, v \in L$ such that $uv \notin L$. While we guess $u$ and $v$, we count the number of symbols guessed, and reject if that number is greater than $2^{2n} + 2^n - 1$.

Now we show the problem is PSPACE-hard. To do so, we observe that $\Delta^*$ is closed, but $\Delta^* \setminus \{w\}$ for $w$ with $|w| \geq 2$ is not. Thus, using our modification of Lemma 3, we could use an algorithm solving the problem of whether a language is closed to solve decidability for polynomial-space bounded Turing machines. ∎

We note that it is possible for an $n$-state NFA $M$ to have the property that $L(M)$ is not closed, but the minimal-length example of a word $uv$ with $u, v \in L$ but $uv \notin L$ is exponentially long. Such an example is given in [4], where it is shown that for some constant $c$, there exist NFA's with of $n$ states such that the smallest word not accepted is of length $> 2^{cn}$.

We note that the problem of deciding, for a given NFA $M$, whether $L(M)$ is open is also PSPACE-complete. The proof is similar to that of Theorem 10.

## 5  Compactness

A closure operator is an *algebraic* closure operator (also called a *finitary* closure operator) if $X^\Box = \bigcup \{Y^\Box : Y \subseteq X \text{ and } Y \text{ finite}\}$. It is easy to show that both Kleene and positive closures are algebraic closure operators. Closed languages form a *complete* lattice under the partial ordering given by set inclusion. The meet and join (infimum and supremum) operators are the following:

(B)  $L \wedge R = L \cap R$.
(A)  $L \vee R = (L \cup R)^\Box$.

A language $L$ is a *compact element* of this lattice if and only if whenever $\{M_i : i \in \mathcal{I}\}$ is a family of languages for some arbitrary index set $\mathcal{I}$ with $L \subseteq \bigvee_{i \in \mathcal{I}} \{M_i\}$, there is some finite $\mathcal{J} \subseteq \mathcal{I}$ such that $L \subseteq \bigvee_{i \in \mathcal{J}} \{M_i\}$. It turns out that our lattice is *compactly generated*, meaning that every language is the supremum of compact elements. It is therefore an *algebraic lattice*, and the compact elements are simply closures of finite languages (as is true for the inclusion lattice of any algebraic closure operator; see [3]). Thus we will say a language is *compact* whenever it is the closure of a finite language. What follows are some results about compact languages.

**Theorem 11.** *Let $L, M \subseteq \Sigma^*$ be compact. Then $(L \cup M)^\Box$ is compact.*

**Theorem 12.** *Let $L \subseteq \Sigma^*$ be compact and let $M \subseteq \Sigma^*$ be finite. Then*

*(1)  $L \cup M$ is compact if and only if it is closed.*
*(2)  $L \setminus M$ is compact if and only if it is closed.*

**Theorem 13.** *Let $L$ be finite and open. Then $L^-$ is compact.*

**Acknowledgment:** This research was supported by the Natural Sciences and Engineering Research Council of Canada.